\documentstyle[11pt,newpasp,twoside,psfig]{article}
\index{isolated neutron stars!RX~J0720.4$-$3125}
\index{isolated neutron stars!RX~J1308.6+2127}
\index{isolated neutron stars!RX~J1605.3+3249}
\index{neutron stars!atmospheres}
\index{neutron stars!magnetic fields}
\index{XMM-Newton!RGS}

\begin{document}
\title{X-ray Spectroscopy of Thermally Emitting Neutron Stars}
\author{M. H. van Kerkwijk}
\affil{Department of Astronomy \& Astrophysics, University of Toronto,\break
60 Saint George Street, Toronto, Ontario, M5S 3H8, Canada}

\begin{abstract}
I describe recent high-resolution X-ray spectroscopy of surface
emission from nearby, thermally emitting neutron stars.  I focus on RX
J0720.4$-$3125, RX J1308.6+2127, and RX J1605.3+3249, all of which
have similar temperature, but differ in the presence and strength of
absorption features in their spectra.  I discuss possible causes for
the absorption we see in two sources, and conclude that it may be
proton cyclotron line absorption, but weakened due to the strong-field
quantum electrodynamics effect of vacuum resonance mode conversion.
\end{abstract}

\section{Introduction}
{\em ROSAT} discovered a number of nearby neutron stars whose emission
appears to be entirely thermal, uncontaminated by accretion or
magnetospheric processes.  At present, six (possibly seven) sources
are known (Treves et al.\ 2000; Haberl 2003).  For four sources,
optical counterparts have been identified.  The high X-ray to optical
flux ratios leave no model but an isolated neutron star.

As a class, the sources are interesting because of the implied
existence of a fair number of neutron stars different from the usual
radio pulsars and X-ray binaries.  At present, their nature remains
unclear.  The original idea, of old neutron stars accreting slowly
from the interstellar medium, has become unlikely because of the very
low accretion rates implied by high proper motions.  Instead, they
might be radio pulsars beamed away from us, although in this case the
long periods are surprising.  Perhaps they have very strong magnetic
fields, and are descendants of anomalous X-ray pulsars and soft
gamma-ray repeaters.

As individuals, the sources are of particular interest because of the
opportunity to study uncontaminated emission from a neutron-star
atmosphere.  The hope is that this will allow one to infer precise
values of the temperature, surface gravity, gravitational redshift,
and magnetic field strength.  In turn, these could be used to
constrain the interior, and hence learn about the equation of state of
cold, ultradense matter, an unexplored region in QCD parameter space.

\section{X-ray Spectra}
Given the interest, long spectroscopic observations were taken with
the {\em Chandra X-ray Observatory} and {\em XMM-Newton}.  The first
results were discouraging: no lines were found in {\em XMM} spectra of
RX J0720.4$-$3125 (Paerels et al.\ 2001), nor in {\em Chandra} spectra
of the prototype of the class, RX J1856.5$-$3754 (Burwitz et al.\
2001), not even after 500~ks (Drake et al.\ 2002; Braje \& Romani
2002).  Instead, for both sources the spectra were found to be
remarkably well described by mildly extincted black bodies.  This was
unexpected, since for light-element atmospheres one would expect a
hard tail (because the opacities decrease towards higher energies),
while for anything else one would expect to see lines (for reviews,
see Pavlov, Zavlin, \& Sanwal 2002; Zavlin \& Pavlov 2002).

\begin{figure}[t!]
\centerline{\psfig{file=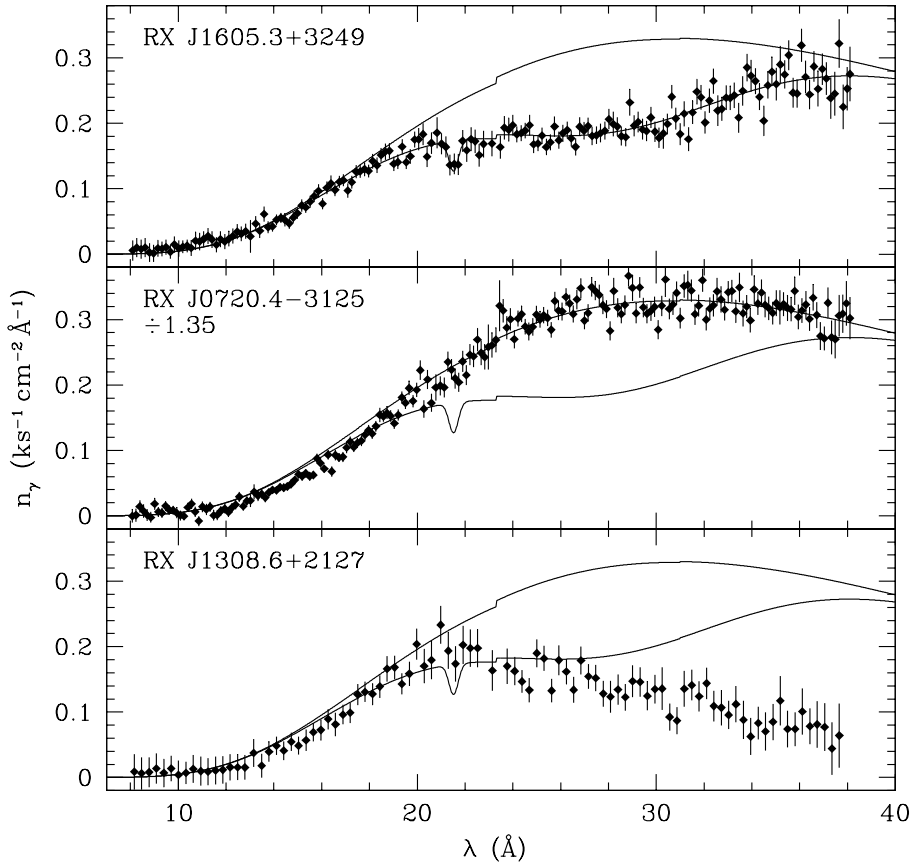,width=0.9\textwidth}~~}
\caption{RGS spectra of three nearby, thermally emitting neutron
  stars.  The top panel shows RX J1605.3+3249, for which the spectrum
  shows clear evidence for an absorption feature.  Overdrawn is the
  best-fit model, a slightly extincted black body with two Gaussian
  absorption features.  The second and third panel show the spectra of
  RX J0720.4$-$3125 and J1308.6+2127, with the same model overdrawn
  for comparison.  Both have similar temperatures, but RX
  J0720.4$-$3125 has no absorption, while RX J1308.6+2127 has much
  stronger absorption.\label{fig:rgs}}
\end{figure}

The situation changed this year.  For RX J1308.6+2127, the
fourth-brightest object in the class, Haberl et al.\ (2003) discovered
a broad absorption feature in {\em XMM} data.  It extends from
$\sim\!0.5\,$keV to lower energies, and can be described by a Gaussian
centred at $\la\!0.3\,$keV.  Furthermore, the third-brightest object,
RX J1605.3+3249, was found to show a weaker, less wide absorption
feature centred at a somewhat higher energy of $0.45\,$keV (Van
Kerkwijk et al.\ 2003), as well as a narrower, marginally significant
absorption line at 0.55\,keV.

In Fig.~\ref{fig:rgs}, the RGS spectra obtained with {\em XMM} for
both sources are shown.  In addition, the spectrum of RX
J0720.4$-$3125 is shown.  This source has a very similar temperature,
$kT\simeq90\,$eV, but a featureless spectrum.

\section{Proton Cyclotron Absorption and Vacuum Resonance}

For RX J1308.3+2127, Haberl et al.\ (2003) suggested that the feature
was due to proton cyclotron absorption.  Given the energy of
$\la\!0.3\,$keV, the implied magnetic field strength is
$\la\!5(1+z)\times10^{13}\,$G (where
$1+z=(1-2GM/Rc^2)^{-1/2}\simeq1.3$ is the gravitational redshift
factor).  This is not unreasonable, given the observed slow spin
period, $P=10.3\,$s.  From the temperature, the neutron star should be
about half a million years old.  Equating this to the characteristic
age, one infers a current spin-down rate $\dot{P}\sim
P/2t\sim3\times10^{-13}$, and a magnetic field of
$3.2\times10^{19}(P\dot{P})^{1/2}\sim6\times10^{13}\,$G, consistent
with what is required.

The large width of the feature, $\sigma_E/E\ga1/3$, is as expected for
proton cyclotron absorption, since the cyclotron energy scales
linearly with the magnetic field strength, which will vary over the
surface (by a factor two for a centred dipole).  Furthermore, as
mentioned by Haberl et al.\ (2003), the equivalent width is consistent
with model calculations of Zane et al.\ (2001).

If we assume the absorption in RX J1605.3+3249 is due to proton
cyclotron absorption as well, we infer
$B\simeq7(1+z)\times10^{13}\,$G.  Unfortunately, no pulsations have
been found -- to a limit of 3\% in the frequency range 0.001--800\,Hz
(Van Kerkwijk et al.\ 2003) -- so we cannot verify this.  Puzzling in
this case, however, is the difference in strength and width of the
features in the two sources.  As mentioned, the temperatures are very
similar, so that cannot be the explanation.

The fields strengths inferred above are in excess of the critical
quantum electrodynamics field $B_{\rm QED}=4.4\times10^{13}\,$G, at
which the electron cyclotron energy equals the electron rest mass.
One effect that may become important is that photons propagating down
the density gradient in the atmosphere can change polarisation mode at
``vacuum resonance,'' where the plasma contribution to the dielectric
properties is compensated by the QED effect of vacuum polarisation.

Vacuum resonance has recently been studied in detail (e.g., Lai \& Ho
2003; see also Lai, these proceedings), and it was found to lead to a
reduction in the contrast of spectral features when it occurs between
the deeper photosphere for the extraordinary mode photons and the
shallower one for the ordinary mode photons.  For the relevant
energies of $\sim\!0.6\,$keV at the surface, this will be the case for
magnetic fields in the range 0.7--$50\times10^{14}$\,G.

If this is correct, it will not be very important for RX J1308.6+2127,
and it makes sense that that source's feature could be reproduced by
Zane et al.'s 2001 models, which do not take vacuum polarisation into
account.  It should affect RX J1605.3+3249, however.  And indeed, it
may explain why the observed feature is so narrow: we might be seeing
only absorption from regions with relatively low field,
$B\la9\times10^{13}\,$G, the contrast of the absorption in regions
with higher field being reduced due to the vacuum resonance.

\section{Future Prospects}

The discovery of absorption features should help further theoretical
work on the vacuum resonance (in progress; Ho \& Lai 2003b).  In
particular, in both sources, the absorption extends up to
$\sim\!0.5\,$keV.  Might it be that the maximum energy out to which
absorption is seen is set by vacuum resonance?  If so, the
observations provide a direct measurement of the critical magnetic
field strength: about $7(1+z)\times10^{13}\,$G.

Theoretical work is also needed on neutral hydrogen, which is strongly
bound in high magnetic fields (for a review, Lai 2001).  Hence, it
should be present, and could lead to observable features.  So far, it
has only been studied at lower fields (e.g., Pavlov \& Meszaros 1993;
Zavlin \& Pavlin 2002).  It might possibly also help understand the
puzzling overall spectral energy distributions.

Observationally, the obvious steps are a survey of the remaining
sources, and using {\em Chandra}/LETGS to obtain spectra over a larger
wavelength range and with phase resolution.  Timing and astrometric
studies could help constrain magnetic field strengths, distances,
space velocities, places of origin, and ages.

All in all, the future looks bright, even if we may learn more about
QED than about QCD.

\acknowledgments I thank Dong Lai and George Pavlov for the good
discussions during the conference, during which the ideas described
here arose.


\begin{references}
\reference {Burwitz}, V., {Zavlin}, V.~E., {Neuh{\" a}user}, R.,
  {Predehl}, P., {Tr{\"u}mper}, J., \& {Brinkman}, A.~C. 2001, \aap,
  379, L35
\reference {Braje}, T.~M. \& {Romani}, R.~W. 2002, \apj, 580, 1043
\reference {Drake}, J.~J. {et~al.} 2002, \apj, 572, 996
\reference {Haberl}, F. 2003, in High Energy Studies of Supernova
  Remnants and Neutron Stars, ed. W.~{Becker} \& W.~{Hermsen}, in
  press [astro-ph/0302540] 
\reference {Haberl}, F., {Schwope}, A.~D., {Hambaryan}, V.,
  {Hasinger}, G., \& {Motch}, C. 2003, \aap, 403, L19
\reference {Ho}, W.~C.~G. \& {Lai}, D.  2003, \apj, submitted
\reference {Lai}, D. 2001, Reviews of Modern Physics, 73, 629
\reference {Lai}, D. \& {Ho}, W.~C.~G. 2003, \apj, 588, 962
\reference {Paerels}, F. {et~al.} 2001, \aap, 365, L298
\reference {Pavlov}, G.~G. \& {Meszaros}, P. 1993, \apj, 416, 752
\reference {Pavlov}, G.~G., {Zavlin}, V.~E., \& {Sanwal}, D. 2002, in Neutron Stars,
  Pulsars, and Supernova Remnants (Garching: MPE), 273
  [astro-ph/0206024] 
\reference {Treves}, A., {Turolla}, R., {Zane}, S., \& {Colpi}, M. 2000, \pasp, 112, 297
\reference {Van Kerkwijk}, M.~H., {Kaplan}, D.~L., {Durant}, M.,
  {Kulkarni}, S.~R., {Paerels}, F.  2003, \apj, submitted
\reference {Zane}, S., {Turolla}, R., {Stella}, L., \& {Treves},
  A. 2001, \apj, 560, 384 
\reference {Zavlin}, V.~E. \& {Pavlov}, G.~G. 2002, in Neutron Stars, Pulsars, and
  Supernova Remnants (Garching: MPE), 263 [astro-ph/0206025]
\end{references}
\end{document}